**MreB helical pitch angle determines cell diameter in *Escherichia coli***

Nikolay Ouzounov[1], Jeffrey Nguyen[2], Benjamin Bratton[1,3], David Jacobowitz[2], Zemer Gitai[1*], Joshua W. Shaevitz[2,3*]

[1]Department of Molecular Biology, Princeton University, Princeton, NJ 08544, USA

[2]Department of Physics, Princeton University, Princeton, NJ 08544, USA

[3]Lewis-Sigler Institute for Integrative Genomics, Princeton University, Princeton, NJ 08544, USA

[*]Corresponding authors: zgitai@princeton.edu, shaevitz@princeton.edu

**Abstract**

Bacteria have remarkably robust cell shape control mechanisms. For example, cell diameter only varies by a few percent across a population. MreB is necessary for establishment and maintenance of rod shape although the mechanism of shape control remains unknown. We perturbed MreB in two complimentary ways to produce steady-state cell diameters over a wide range, from 790±30 nm to 1700±20 nm. To determine which properties of MreB are important for diameter control, we correlated structural characteristics of fluorescently-tagged MreB polymers with cell diameter by simultaneously analyzing 3-dimensional images of MreB and cell shape. Our results indicate that the pitch angle of MreB inversely correlates with cell diameter. Other correlations are not found to be significant. These results demonstrate that the physical properties of MreB filaments are important for shape control and support a model in which MreB dictates cell diameter and organizes cell wall growth to produce a chiral cell wall.

**Introduction**

Bacteria come in a large variety of shapes and sizes. Their diameters can range from 200 nm in the case of *Mycoplasma* to 750µm or more for the macroscopic *Thiomargarita namibiensis* (1). Cells also come in a variety of shapes from spheres and rods to spirals and squares. These shapes can be important for many aspects of bacterial life such as motility, growth, predation, and packing within biofilms (2).

Cell shape in the majority of Gram-positive and Gram-negative bacteria is defined by the rigid, exoskeletal peptidoglycan (PG) cell wall (3). Bacterial cells modify their existing cell wall by cutting existing PG and inserting new glycan strands during growth and division (4). In most rod-shaped bacteria, such as *Escherichia coli* and *Bacillus subtilis*, the bacterial actin-homolog MreB organizes the cell-wall growth machinery such that cell wall insertion is spatially driven by the localization of MreB (5-7). MreB is required for rod shape, and thus *mreB* deletions or treatment with MreB polymerization-inhibiting drugs such as A22 lead to spherical cells with weakened cell walls that are prone to lysis (5, 8).

The structural properties of MreB and MreB polymers have only recently begun to be elucidated. *In-vitro* experiments suggest that MreB forms anti-parallel filaments that bind to membranes (9). These filaments induce curvature similar to that found in cells when bound to the outside of membrane vesicles. The length of MreB filaments is a heavily debated subject. Estimates of *in-vivo* MreB polymer lengths range from below the detection limit (10) to over 1.4 um (11). Swiulius et al. concluded that filaments were shorter than 80nm based on the fact that unlabeled MreB polymers have not been observed in cryo-electron microscopy (cryo-EM), although the sensitivity to detect thin polymers near a membrane has not been demonstrated with this technology. Reimold et al. estimated the lengths of filaments in cells expressing MreB-mCherry to be micron-sized using structured illumination microscopy (SIM) (12), but they were unable to gather 3D data due to the poor quantum efficiency of mCherry. To date no studies have demonstrated that polymer length or any other feature of MreB polymers is important for cell shape determination.

Recently, it was shown that fluorescent protein fusions can generate protein aggregation depending on the fluorescent protein used, resulting in misfolding or mislocalization of the protein (13). This is a possible cause for the shape anomalies observed in cells expressing MreB-mCherry and could affect the observed MreB structure. More representative *in-vivo* measurements of MreB polymers should be attainable through the use of better fluorescent proteins and a 3-dimensional polymer measurement technique.

We previously confirmed the long-held hypothesis that MreB spatially localizes the various synthase molecules responsible for modifying and inserting material into the peptidoglycan cell wall (14). However, two competing hypotheses link MreB localization to growth (15, 16). MreB has been suggested to serve as a scaffold that co-localizes multiple synthase molecules at random positions along the cell cylinder (6). Other evidence suggests that MreB polymers serve to spatially organize cell-wall insertion over long distances in a helical pattern that gives rise to the growth of an ordered cell wall and twisting (17, 18). Given these conflicting views, a key question that has emerged is whether spatial organization of MreB is important for cell shape regulation. Furthermore, MreB structures have many different properties such that if MreB organization is important for shape control, we must understand whether that is a result of the amount of MreB, the length of the polymers, the orientation of the polymers, or other properties of MreB.

Here we address these key outstanding questions by generating a better MreB fluorescent fusion and using high-resolution, 3D imaging to quantitatively correlate MreB's physical properties with cell diameter across a range of MreB perturbations that also perturb shape. We show that the only property of MreB that significantly correlates with cell diameter is the helical pitch angle of MreB filaments. These results provide the first evidence that the structure and organization of MreB filaments is important for defining cell shape. Our findings also support a model for cell shape determination where the helical conformation of MreB polymers gives rise to helical PG insertion, which in turn leads to different cell diameters due to changes in the organization of the cell wall.

## Results and Discussion
### MreB$^{msfGFP}$ is minimally perturbative

Previous live-cell fluorescence microscopy studies of MreB localization used fluorescent fusions of YFP to the N-terminus of MreB or of mCherry inserted internally at a non-conserved surface-exposed loop in the protein (12). Both of these strains suffered from physiological defects. The N-terminal YFP fusion, which manifests as a single large helical filament, does not complement a deletion of the MreB protein. On the other hand, the mCherry internal "sandwich" fusion yields multiple small fluorescence structures and does rescue the viability of MreB deletion strains, but frequently results in significant cell shape defects suggesting that it too disrupts MreB function (Fig. S1).

Because mCherry has recently been shown to stimulate aggregation when fused to proteins (13), we sought to find a better probe for live-cell studies by replacing the mCherry in the sandwich fusion with 9 different fluorescent proteins, 6 of which have been shown to cause the least amount of aggregation (Table S1, 13). The majority of the fluorescent proteins tested were not able to restore rod shape in place of the sandwich fusion (Fig. S1). The fusion that generated the most native cell shape was MreB$^{msfGFP}$ (monomeric-super-folder-GFP) encoded on the chromosome at the native *mreB* locus under native regulation.

We used two criteria to determine the level of perturbation that tagging MreB with msfGFP would cause. First, the tagged strain should have the same growth rate as the untagged version when grown in different media. We grew the fluorescently tagged and untagged strains in three kinds of media: high-sucrose media (M-media), rich Luria Broth (LB), and a minimal media with glucose and casamino acids (M63). MreB$^{msfGFP}$ shows unperturbed growth rate in all media when compared to the unlabeled wild type (Fig. S2). Both strains exhibited identical exponential growth doubling times in all media, with doubling times of 19 minutes in M-media, 17 minutes in LB, and 28 minutes in M63. Our second criterion was that the fluorescent fusion should minimally alter cell width as we were primarily interested in studying MreB's role in determining cell shape. Using our custom cell-shape analysis software (see Materials and Methods), we found that MreB$^{msfGFP}$ is approximately 5% wider than the unlabeled parental strain and equally as rod-shaped (Fig. S3). The average diameters of the unlabeled and labeled cells are 893±3 nm and 934±6 nm respectively (All values are reported as mean ± 80% confidence interval). Using these two criteria, we conclude that the MreB$^{msfGFP}$ fluorescent fusion is minimally perturbative.

**MreB$^{msfGFP}$ forms ~500-nm long polymers along the cell cylinder**

We next examined MreB$^{msfGFP}$ and calculated its polymeric properties. Combining 3D imaging with polymer detection and segmentation software developed in our lab, we were able to calculate the size and orientation of MreB structures with respect to the 3D cell surface. We used a previously-developed forward convolution fitting method to estimate the 3D cell shape of cells stained with FM4-64 (14). Polymer detection was performed by fitting the 3D fluorescent images of MreB$^{msfGFP}$ with linear segments confined to lie on the measured surface of the cell (Fig. 1A, Materials and Methods). This analysis revealed that MreB forms extended structures that are larger than the diffraction limit. In M63 media, we measured the average polymer length of MreB$^{msfGFP}$ to be 500±10 nm(Fig. 2B). Examples of fits to both MreB$^{msfGFP}$ structure and cell shape are shown in Fig. 1B, with the surface color showing the fluorescence intensity of MreB$^{msfGFP}$ at each point on the surface, and the detected polymers shown in white. We measured on average 7 polymers per cell and these polymers were mostly found in the cylindrical portion of the cell and excluded from the cell poles (Fig. S4).

Our analysis also enabled us to determine the helical pitch angle of MreB polymers relative to the cell long axis. Due to the effects of blurring by the microscope, pitch angle was only calculated for polymers with a length greater than 300 nm. On average, MreB$^{msfGFP}$ polymers had a right-handed helical pitch angle of 91±1°. The angle is measured relative to the long axis of the cell (Fig. 1C), with angles above 90° indicating a right-handed pitch (p=0.08). Interestingly, the handedness reported is opposite of the left-handed polymers we previously measured for *E. coli* using the non-complimenting N-terminal YFP fusion. Based on the functionality of this fusion, we have a higher confidence that this measurement more faithfully represents the normally unlabeled state of MreB and can be used as a basis for further study of MreB function.

**Generating different cell shapes by perturbing MreB**

Armed with tools for the 3D quantification of cell shape and MreB polymer conformation in the wild-type MreB$^{msfGFP}$ strain, we sought to probe changes in bacterial cell shape due to the perturbation of MreB. Amino acid substitutions and the application of the MreB polymerization-inhibiting drug A22 at sub-lethal concentrations both result in cell shape changes that are directly linked to MreB. Although these treatments have been previously shown to cause changes in cell shape, no studies have examined the biophysical properties of MreB polymers to determine the cause of the observed cell shape modification.

A22 has been shown to inhibit rod shape through a specific interaction with MreB (8, 19). Growing *E. coli* in the presence of different sub-lethal concentrations of A22 resulted in the growth of cells with varying steady-state cell diameters (Fig. 3A). We used concentrations of A22 up to 1 µg/mL because higher concentrations lead to high lethality and potential disruption of nonspecific targets (20). Using this range of A22, we reproducibly generated cell populations with stable steady-state diameters ranging from 934±6 nm without the drug to 1700±20 nm at 1 µg/mL. Each of these treatment conditions contain measurements ranging from 42 to 474 cells, with an average of 297 cells.

To generate MreB-dependent changes in cell shape genetically, we created a collection of mutations in $mreB^{msfGFP}$. The simplest way to generate such a collection of mutations is by selecting for A22-resistant suppressor mutations. Using this method, we isolated 12 different MreB amino-acid substitutions in $MreB^{msfGFP}$ which showed varying levels of A22 resistance. Importantly, the mutations were distributed across 3 of MreB's 4 subdomains (IA, IB, and IIA, Fig. S5), indicating that the mutations likely perturb a range of MreB properties. We also used site-directed mutagenesis to generate A53T and the deletion of the 53$^{rd}$ amino acid (ΔA53) as this residue was previously shown to increase cell diameter depending on the amino acid substitution (21). We grew all strains in M63 media with casamino acids and measured between 64 and 1228 cells (478 on average) for each of the 14 different mutations. These mutations altered the average cell diameter compared to the wild type $MreB^{msfGFP}$ during steady state growth (Fig. 3B) generating cells ranging in diameter from 790±30 nm (thinner than wild type) to 1590±60 nm (wider than wild-type). Depending on the specific amino acid substitution, the average MreB polymer length ranged between 328±7 nm to 690±50 nm (Fig. 4A).

**MreB polymers in A22-treated cells and the effect of A22 on suppressor mutants**

While high concentrations of A22 abolish MreB polymers (19, 22), the effect of sub-lethal concentrations on MreB structure is not known. As expected, we find that MreB polymer length is inversely correlated with the concentration of A22 in wild type cells when moderate amounts of the drug are used (Fig. S6). Lengths ranged from 500±10 nm for untreated cells to 440±10 nm for cells at our maximum treatment level of 1 μg/ml. This result is consistent with previous work showing that MreB bound to A22 weakens the inter-protofilament interaction (8, 19), decreasing the stability of MreB polymers.

We next examined A22 resistance and polymer length in each mutant. A22 resistance increased with polymer length, likely due to specific amino acid substitutions leading to polymer stabilization and a reduction in the turnover of MreB monomers (Fig. 4B). It has been hypothesized that the ADP to A22 exchange happens in the post-ATP-hydrolysis monomeric state (23). This would stabilize MreB states that promote an ATP-bound conformation and would increase A22 resistance. The most A22-resistant mutant we isolated was E143A, which was previously proposed to be deficient for ATP hydrolysis (22). The cell diameter for this strain is 930±40 nm, identical to that of the unmutated strain, and it shows robust growth in A22 concentrations less than 100 µg/ml. Above this level, A22 begins to bind non-specifically to the ATP pocket of other proteins which increases lethality (20). Future work on the MreB$^{E143A}$ mutant could shed light on MreB function as it is currently thought that the turnover of MreB monomers is physiologically necessary.

**Correlation analysis between MreB polymeric properties and cell morphology**

Our fluorescence analysis of dual-labeled cells yields a number of quantitative metrics of cell shape and MreB polymer conformation. These include cell width, cell length, cell volume, polymer number, polymer size, polymer helical pitch angle relative to the cell center line, and the fraction of MreB that appears to reside on the inner membrane. We used a correlation analysis of these quantitative metrics to investigate which properties of MreB were most predictive of changes in cellular morphology (Fig. 5).

The two strongest correlations with cell diameter in both data sets are MreB helical pitch angle and MreB polymer number. In order to analyze the significance of correlations that are consistent between the two datasets, we combined the data from both treatment cases and recalculated the correlation (Fig. 4C). The largest correlations were between cell diameter and MreB polymer angle (-0.69, p=0.023) and between cell diameter and polymer number (0.76, p<0.001). Both show significance p<0.05 after accounting for the effect of multiple comparisons.

**MreB pitch angle is highly predictive of cell diameter for both A22 and mutant experiments**

The correlation between MreB polymer number and cell diameter is due to the coupling between cell volume and number of MreB proteins. Cell volume is strongly correlated with cell diameter and cells with more volume, and hence more total protein including MreB, create more detectable polymers. This results in a significant correlation between polymer number and diameter with correlation coefficients of 0.70 and 0.92 for the mutant and A22 treatment data sets respectively. We also observed a reduction in the total number of MreB polymers as polymer length increased, with an inverse correlation coefficient value of -0.71 and -0.64 for the mutant and A22 treatment data sets respectively. This is expected as a larger percentage of MreB is present in the longer polymers, decreasing the total number of short polymers. Thus, polymer length and number correlations confirm previous findings that MreB assembly is important but that these properties are not specifically informative with respect to cell diameter control.

In contrast, the unexpected inverse correlation between MreB pitch angle and cell width across all conditions tested yields new insight into shape control. We observed an inverse correlation between MreB helical pitch angle and cell diameter, with correlation coefficients of -0.78 and -0.95 for the mutant and A22 treatment data sets respectively. The sign of this correlation indicates that a reduction in pitch angle leads to an increase in the cell width. Interestingly the average handedness of the polymers changed from the right-handed (93±1°) to left-handed (84±3°) crossing 90° as cells get wider. This observation is in agreement with previous work showing that the growth twist of *E. coli* switches handedness with increasing concentration of A22 (18)

**A22-specific correlations**

We also observed correlations specific to the A22 treatment. We measured an inverse correlation coefficient of -0.57 between MreB polymer length and cell diameter in the A22 data treatment, but the same correlation was insignificant in the mutants (See Materials and Methods for significance determination). The inconsistencies between the correlations from A22 treatment and the mutants are likely due to the mechanism by which A22 effects MreB polymers, namely reducing the concentration of MreB monomers available for polymerization. In the mutant data set, we see that mutants with longer polymers are more resistant to A22 with a correlation coefficient of 0.62. Mutations that lead to longer polymers stabilize MreB in its polymeric form and counteract the effects of A22.

The correlation between the fraction of MreB that is membrane bound and cell diameter shows opposite correlations in the two data sets. For the A22 treatment, there is an increase in the cytoplasmic portion of MreB as cell diameter increases, with a correlation coefficient of -0.80. This is consistent with previous work that shows A22 increases the diffuse MreB monomer pool by preventing polymerization (8). In the MreB mutant data set, the reverse trend is seen and the membrane bound MreB fraction is positively correlated with cell diameter with a value of 0.69. This contradiction suggests that the proportion of MreB localized on the membrane is not a direct contributor to the determination of cell diameter.

**A proposed mechanism for MreB-mediated helical growth**

Prior to this study, we hypothesized that the helical pitch angle of MreB could help organize cell wall synthesis in a manner that leads to chiral growth twist. Our simulations suggested that the chiral order of the peptidoglycan would primarily alter cell diameter. At that time, we had no way of making measurements of functional polymers, nor a way to alter the helical pitch angle of MreB to test this hypothesis. Here we addressed both of those limitations by generating a minimally-perturbative MreB fluorescent fusion, using 3D imaging and automated image analysis to quantify both polymer and cell shape characteristics, and using both A22 treatment and mutagenesis to alter MreB. Our analysis of correlations between all biophysical parameters showed that cell diameter has a significant correlation with polymer angle regardless of treatment with a correlation coefficient of -0.95 in the A22 data set and -0.78 in the mutants. Both treatments lead to similar cell diameters for a measured angle (Fig. 4C), and a significant correlation is observed across datasets. The similarity across these independently-derived sets of MreB perturbation supports the conclusion that helical pitch is a key determinant of cell diameter.

We hypothesize that the mutations and A22 treatment alter the mechanics of the interaction between neighboring MreB monomers and/or between the MreB polymers and the cell membrane, resulting in a change in the angle of the helical fragments relative to the curvature of the cylindrical cell (24). In cells treated with A22, MreB mimics its ADP bound state (23). A22 treatment and mutations perturbing the nucleotide-binding pocket of MreB have the potential to alter the geometry of the polymer as it has been shown that the angle between adjacent MreB monomers depends on the state of the bound nucleotide (25). In addition, mutations in the MreB-MreB binding surface or near the membrane-binding domain can directly affect the structure of the polymer and higher-order filaments. These types of structural changes have been shown to affect key parameters in the determination of the helical conformation of MreB, such as the stiffness, bending angles, and twisting angles (24, 26), which could result in polymers with different helical pitch angles inside the cell.

Changes in helical pitch angle can affect the chiral organization of the peptidoglycan cell wall (17). It has been recently shown that as cells are treated with increasing amounts of A22, the chirality of their cell wall shifts from right handed to left handed. Thus, changes in the MreB structure appear to propagate to the structure of the cell wall (18). Moreover, we observe not only a change in the degree of the helical pitch angle, but also that the angles cross $90^{\circ}$, indicating a change in handedness of the MreB. This consistency between our observed change in handedness and the change in handedness of the MreB polymer motion and the cell growth twist recently observed by Tropini et al. supports our model (18).

**Conclusion**

MreB was previously shown to be necessary for establishment and maintenance of rod shape in its role localizing the cell wall insertion machinery. However, the mechanism via which rod-shaped bacteria establish specific diameters remained unclear. Bacteria such as *E. coli* and *B. subtilis* exhibit a chiral growth twist that is determined by the organization of the PG cell wall. Here, we examined the biophysical properties of MreB polymers using 3D fluorescence microscopy and a minimally inhibitory MreB fluorescent fusion. MreB mutations or treatment with sub-lethal concentrations of A22 both lead to cell shape changes and changes to the biophysical properties of MreB polymers. We detected a consistent correlation between MreB helical pitch angle and cell diameter, suggesting that a major role of MreB is to set the cell diameter by organizing a chiral cell wall structure. Importantly, these results indicate that rather than simply acting as a scaffold to cluster various aspects of the cell growth machinery, proper cell shape control requires MreB to form extended polymers and that the biophysical properties of MreB polymers can dictate specific aspects of morphology such as cell diameter.

These results are consistent with our previous hypothesis that the helical pitch angle of MreB dictates the chiral organization of the PG cell wall, and they allow us to build a model that links the atomic-scale properties of MreB to the micron-scale cell shape. Specifically, we propose that the MreB molecular structure determines the intrinsic morphology of the polymer and the helical localization pattern of MreB on the cell membrane. This patterning regulates the chiral organization of the cell wall, which in turn, contributes to the control of diameter of the cell as new material is inserted heterogeneously into the pressure-strained cell wall. This work brings new insights into the multiple functions of MreB. Not only is MreB essential for establishment and maintenance of rod shape, but it helps establish the specific diameter of cells at steady-state growth.

Though we have shown one role for MreB in determining steady-state diameter, bacteria can also dynamically adjust their morphology in response to environmental and internal conditions. The final cell diameter is likely a complicated function of the MreB polymer angle, cell turgor pressure, and details of PG insertion governed by growth rate and nutrient availability. It will be interesting to see if there are similar principles connecting MreB and cell shape if width is altered by changing other parameters such as pressure or growth rate. In addition to MreB, there are a number of MreB-associating proteins. How these proteins interact with MreB and influence its polymeric structure remains unclear. The division machinery also likely plays a role in the determination of cell diameter (27), and it may be possible to investigate this by performing experiments similar to those described here with the tubulin homolog FtsZ.

**Materials and methods**

*Construction of MreB$^{msfGFP}$*

The construction of MreB$^{msfGFP}$ was previously described (14). The specific MG1655 strain differs from that precous work as we found that there are physiological and metabolic differences between MG1655 strains in different labs, presumably from accumulation of genomic mutations over time. For this reason, we chose to use MG1655 that could be traced to back to the Yale Coli Genetic Stock Center. We moved the *csrD-kanR-mreB$^{msfGFP}$-mreCD* operon from our previous MG1655 to MG1655 (CGSC #7740) using the lambda red method followed by selection for kanamycin resistance (28). Colonies were picked and screened using fluorescence microscopy and then sequenced.

*Media conditions*

Multiple media compositions are used for comparison of cell shape between fluorescently labeled and unlabeled MreB strains. The three media used are M media, Lysogeny broth with 5g NaCl per liter and M63 with glucose and casamino acids (29, 30). All measurements of MreB polymers and cell shape were conducted in M63 media. Kanamycin sulfate (sigma) at 20 ug/mL was used in overnight cultures and plating, but was not used in subcultures used for imaging.

*Selection of mutants*

Individual colonies of MreB$^{msfGFP}$ were grown overnight. Cultures were spread the following day on LB plates containing 30 µg/mL kanamycin, 1.5 µg/mL cephalexin, and 10 µg to 35 µg A22. We find that 1.5 µg/mL cephalexin aids in the selection of A22 suppressors since cells that lose MreB function are hypersensitive to cephalexin. Individual colonies that grow on the plates were grown in liquid for imaging and mreB was sequenced to identify mutations.

*A22 resistance quantification*

Each strain was grown in a 96 well plate in LB containing serial dilution of A22 ranging from 100 to 0 µg A22. We used a BioTek™ microplate reader to record the OD600 over 16 hours of growth at 37$^o$C. The A22 resistance of each MreB mutant was calculated by its IC50.

*Imaging of mreB mutants*

Strains were grown over night at 37$^o$C in M63 media in the presence of Kanamycin in order to prevent contamination. The next morning, cultures were subcultured in the morning to 1:10000 to 1:30000. When OD600 of the culture is between 0.15 to 0.3 cells are imaged on 1%

UltraPure™ Agarose (life technologies) pads made of M63 media. Imaging was conducted in a 20°C temperature-controlled room on a custom-built inverted wide field fluorescent microscope with a 1.43NA 100x objective (Nikon). Images for stacks were taken at 100 nm increments in stage position.

*A22 treatment imaging*

MreB$^{msfGFP}$ strain was grown over night at 37°C in the presence of Kanamycin in order to prevent contamination. The next morning, cultures were subcultured to 1:10000-1:30000 in the presence of A22 at concentrations of 0 μg, 0.125 μg, 0.25 μg, 0.5ug, 0.75 μg and 1 μg A22. When OD600 of the culture was between 0.15 to 0.3, cells were imaged on 1% UltraPure™ Agarose (life technologies) pads made of M63 media containing the same concentration of A22 as the liquid media. Imaging was then conducted in the same manner as with the *mreB* mutants.

*3D cell shape reconstruction*

Cell shapes were measured by fitting 3-dimensional images of cells stained with FM4-64 with an active mesh in MATLAB (Mathworks). An initial surface is found by fitting a series of active contours to axial slices of the cell (31). Convolving the surface with the 3-dimensional point-spread function (PSF) of our microscope creates a test image that can be compared directly to the image stack from the microscope. The surface is then iteratively deformed to minimize the square difference between the simulated image and the real image (32). The PSF was measured by averaging image stacks of multiple individual 0.1-μm TetraSpeck™ microspheres imaged at 100 nm steps in the axial direction.

*MreB polymer measurements*

Polymers were measured by fitting the 3-dimensional MreB images to a set of polymers confined to the membrane. Using the surface determined by the membrane fitting, a 2-dimensional unwrapped image of the MreB polymers is created. Segmentation is performed on this image to determine the number, location, and initial length of the polymers on the surface. Each polymer is modeled by an active contour confined to the membrane and then deformed to fit the 3D MreB image. The polymer positions are convolved with the PSF to create a simulated MreB image. Once again, the polymers deform to minimize the square difference between the simulated image and the image from the microscope. The length of the polymer is measured in 3D and the polymer angle is measured as the angle away from the circumferential direction, with positive angles corresponding to a right handed helical pitch and negative angles corresponding to a left handed helix.

The surface is then used to calculate measurements of cell shape such as length, volume and diameter. Radius is measured as the average distance from the surface to the centerline after the removal of the pole regions. Membrane fraction is measured by comparing the MreB image to that of the FM4-64 image and a simulated image in which the cell envelope is filled with uniformly distributed fluorophore. After the poles are removed, axial projections along the centerline of each of the images are taken and then a radial projection is performed on this. The MreB projected image is represented as a linear combination of the normalized FM4-64 membrane projection and the filled simulation. The fraction of MreB on the membrane is the weighting term of the membrane projection.

*Correlation significance testing*

Pearson correlation coefficients were used to analyze relationships between different datasets. In order to determine if correlations between two parameters were significant, we compared our data to a noise model in which both datasets were shuffled 10,000 times to create a distribution of correlation values. Correlations were significant if the values were within the 99.67$^{th}$ percentile of the noise distribution, which is equivalent to p <0.05 after accounting for multiple hypothesis testing using the Bonferroni Correction with 15 tests.

**Acknowledgments**

This work was supported by awards from the National Science Foundation (PHY-0844466, PHY-1022140) and the National Institutes of Health (R01GM107384, P50GM071508). The authors thank Thomas Silhavy, Kerwyn Huang, Tristan Ursell, Alexandre Colavin, and Grant Jensen for helpful discussions.

**Figure Captions**

Figure 1. Cell shape and polymer fitting. (A) Diagram outlining the cell shape and polymer fitting algorithm. Cells expressing MreB$^{msfGFP}$ under native regulation are membrane stained with FM4-64 and imaged using fluorescent microscopy. The imaging process can be written as a convolution between the point-spread function (PSF) of the microscope and the spatial distribution of fluorescent molecules, in this case the membrane. To estimate the shape of the surface, a model cell is convolved with the PSF to create a simulated image. The surface is relaxed so that the simulated image best matches the experimental image. A similar process is used to fit the MreB polymers. An initial segmentation is performed to estimate the initial number of polymers. Each polymer is modeled as a stiff elastic rod confined to the surface of the membrane. Again, a simulated MreB image is created and model filaments relax to best match the experimental image. The angle of the MreB filaments are measured from as the average angle of the filament relative to the axial direction of the cell. (B) Representative surface fits of cells expressing MreB$^{msfGFP}$. The surfaces are fit from 3-dimensional images of FM4-64 stained cells. The color of the surface is determined by interpolating the intensity of the 3D MreB image at the points of the surface. The detected polymers are shown in white. (C) Figure showing the measurement of angle of the polymers. After fitting the polymers to the 3D image, the polymer angle is measured in an unwrapped space as the angle the polymer makes away from the cylindrical axis of the cell. Angles larger than 90° indicate a right-handed helical pitch.

Figure 2. Probability density functions of (A) cell diameter, (B) MreB polymer lengths, and (C) MreB polymer angles in *E. coli* expressing MreB$^{msfGFP}$. Data is derived from 459 cells, with an average of 7.3 polymers detected per cell. The distribution in (A) shows an average diameter of 934±6. The average MreB polymer length was 500±10 nm . The angle distribution of monomers in (C) is made by weighting the angle distribution of polymers by the length of each polymer, and had a mean angle of of 91±1°. All values are shown as mean ± 80% CI.

Figure 3. Two independent methods to perturb cell diameter. (A) Cells expressing MreB$^{msfGFP}$ were grown to steady state in different concentrations of the MreB-polymerization-inhibiting drug A22. All treatments are below the lethal concentrations of A22. As A22 concentration increases, cells significantly increase their diameter. (B) Cell diameter can also be changed with single point mutants in the *mreB$^{msfGFP}$*. 14 mutants were generated with diameters both larger and smaller than the unmutated form (WT). Mutants are arranged in order of increasing diameter. Error bars represent 80% confidence intervals.

Figure 4. MreB polymer measurements. (A) MreB mutations can alter the MreB polymer length. In some cases the polymer length is increased with respect of wild type and in other cases the polymer length is decreased. Mutations are arranged as in Figure 2A, in order of increasing average cell diameter. Error bars represent 80% confidence intervals. (B) MreB polymer length correlates with the level of A22 resistance of the MreB mutants. Mutations that have longer polymers have high levels of A22 resistance while increased sensitivity to A22 is seen in mutations that have shorter polymers. A22 resistance levels are shown as a fold changes from wild-type MreB$^{msfGFP}$, with levels capped at 100-times that of wild-type since higher levels of A22 can effect proteins other than MreB. (C) The average polymer angle inversely correlates with the cell diameter in both mreB mutant data (Black) and A22 treatment data (Green). The cell diameter increases as the average helical pitch angle of MreB decreases. The handedness of the helical pitch changes in both data sets as 90º is crossed. Unperturbed MreB$^{msfGFP}$ is indicated by the A22 treatment with the smallest diameter.

Figure 5. Cell shape was modified using either point mutants in *mreB$^{msfGFP}$* or treatment with sublethal concentrations of A22. Correlation maps between cell shape metrics and measured MreB properties were created for each set of conditions: (A) Mutants (15 conditions), (B) A22

treatment (6 conditions) and (C) combined data sets (19 conditions). † denotes p=.023 and ‡ indicates p<.001, after accounting for multiple comparison.

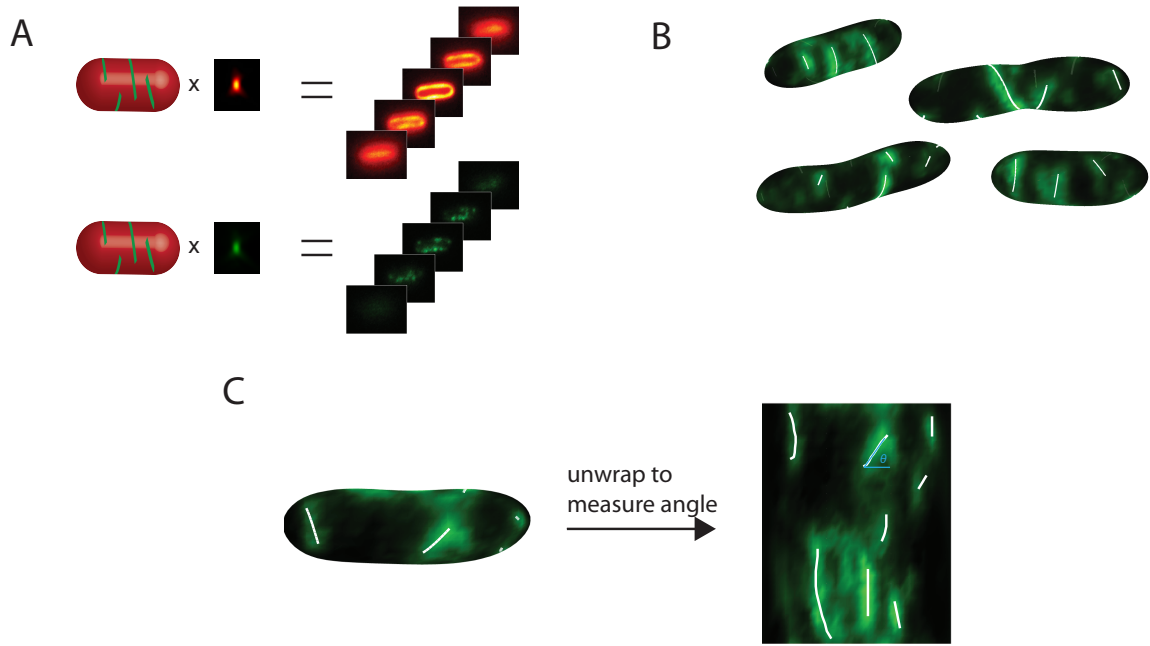

Figure 1

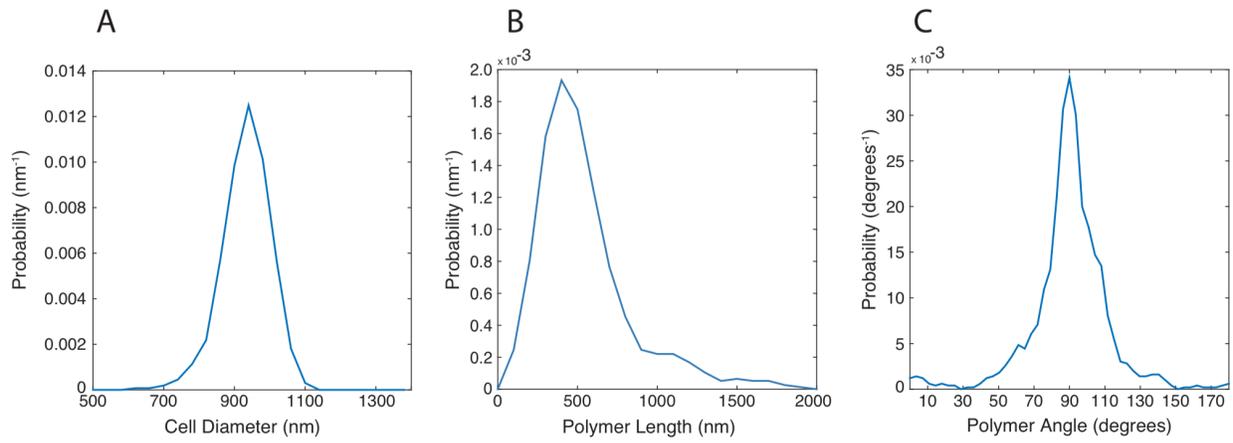

Figure 2

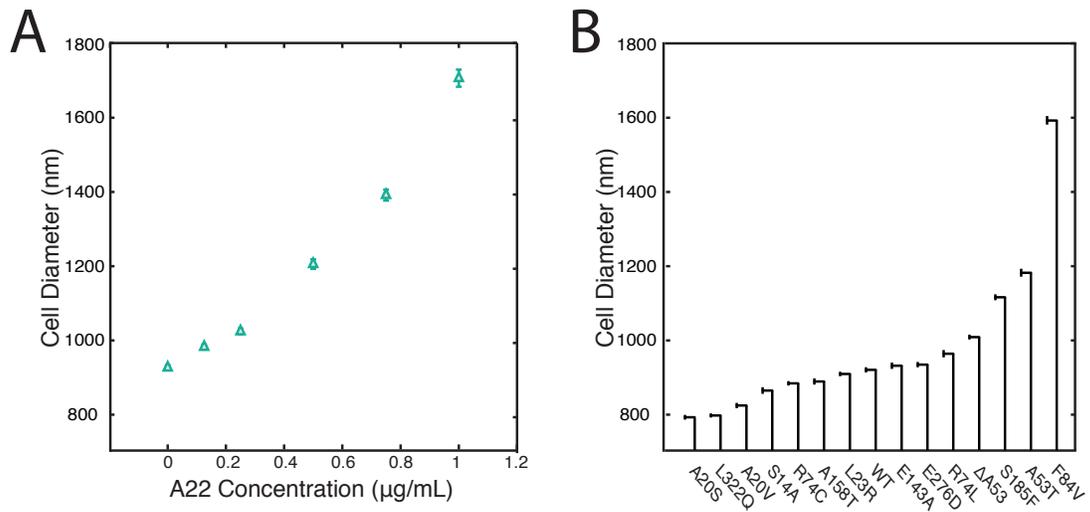

Figure 3

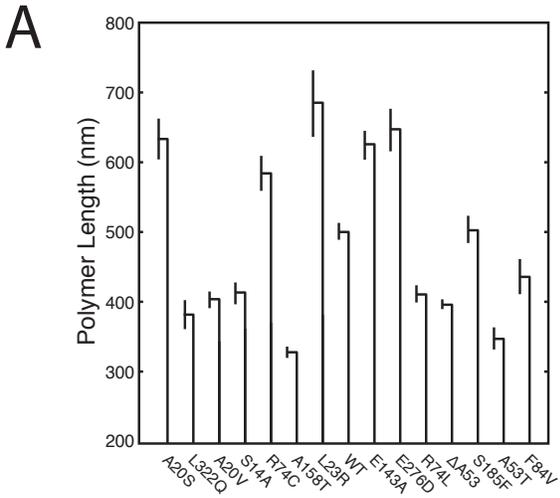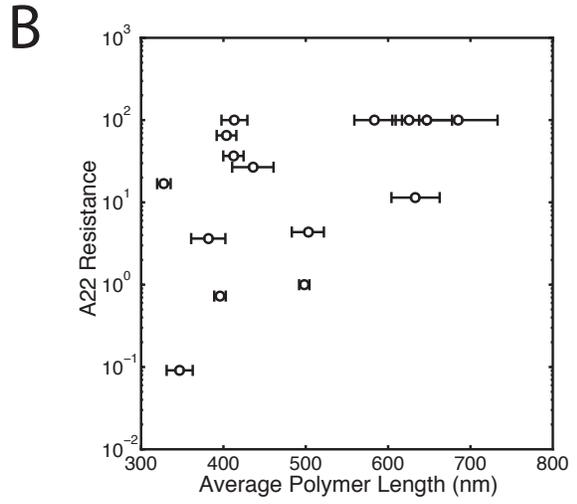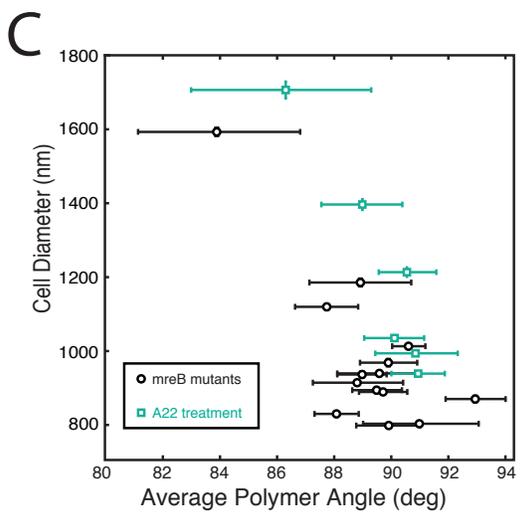

Figure 4

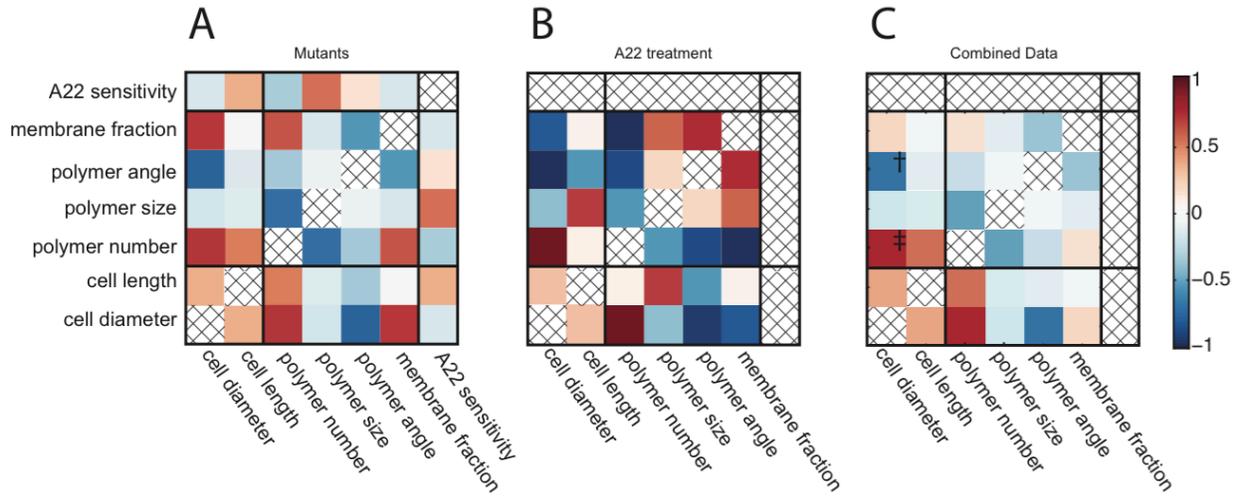

Figure 5

# Supplemental Materials for MreB helical pitch angle determines cell diameter in *Escherichia coli*

Nikolay Ouzounov[1], Jeffrey Nguyen[2], Benjamin Bratton[1,3], David Jacobowitz[2], Zemer Gitai[1*], Joshua W. Shaevitz[2,3*]

[1]Department of Molecular Biology, Princeton University, Princeton, NJ 08544, USA

[2]Department of Physics, Princeton University, Princeton, NJ 08544, USA

[3]Lewis-Sigler Institute for Integrative Genomics, Princeton University, Princeton, NJ 08544, USA

| | |
|---|---|
| MreB | MLKKFRGMFSNDLSIDLGTANTLIYVKGQGIVLNEPSVVAIRQDRAGSPKSVAAVGHDAKQMLGRTPGNIAAIRPMKDGVIADFFVTEKMLQHFIKQVHSNSFMRPSPRVLVCVPVGATQVERRAIRESAQGAGAREVFLIEEPMAAAIGAGLPVSEATGSMVVDIGGGTTEVAVISLNGVVYSSSVRIGGDRFDEAIINYVRRNYGSLIGEATAERIKHEIGSAYPG==SGSS====xxxxx====SGAPG==DEVREIEVRGRNLAEGVPRGFTLNSNEILEALQEPLTGIVSAVMVALEQCPPELASDISERGMVLTGGGALLRNLDRLLMEETGIPVVVAEDPLTCVARGGGKALEMIDMHGGDLFSEE |
| mCherry | MVSKGEEDNMAIIKEFMRFKVHMEGSVNGHEFEIEGEGEGRPYEGTQTAKLKVTKGGPLPFAWDILSPQFMYGSKAYVKHPADIPDYLKLSFPEGFKWERVMNFEDGGVVTVTQDSSLQDGEFIYKVKLRGTNFPSDGPVMQKKTMGWEASSERMYPEDGALKGEIKQRLKLKDGGHYDAEVKTTYKAKKPVQLPGAYNVNIKLDITSHNEDYTIVEQYERAEGRHSTGGMDELYK |
| msfGFP | SKGEELFTGVVPILVELDGDVNGHKFSVRGEGEGDATNGKLTLKFICTTGKLPVPWPTLVTTLTYGVQCFSRYPDHMKQHDFFKSAMPEGYVQERTISFKDDGTYKTRAEVKFEGDTLVNRIELKGIDFKEDGNILGHKLEYNFNSHNVYITADKQKNGIKANFKIRHNVEDGSVQLADHYQQNTPIGDGPVLLPDNHYLSTQSKLSKDPNEKRDHMVLLEFVTAAGITHGMDELYK |
| mVenus | MVSKGEELFTGVVPILVELDGDVNGHKFSVSGEGEGDATYGKLTLKLICTTGKLPVPWPTLVTTLGYGLQCFARYPDHMKQHDFFKSAMPEGYVQERTIFFKDDGNYKTRAEVKFEGDTLVNRIELKGIDFKEDGNILGHKLEYNYNSHNVYITADKQKNGIKANFKIRHNIEDGGVQLADHYQQNTPIGDGPVLLPDNHYLSYQSKLSKDPNEKRDHMVLLEFVTAAGITLGMDELYK |
| Venus | MVSKGEELFTGVVPILVELDGDVNGHKFSVSGEGEGDATYGKLTLKLICTTGKLPVPWPTLVTTLGYGLQCFARYPDHMKQHDFFKSAMPEGYVQERTIFFKDDGNYKTRAEVKFEGDTLVNRIELKGIDFKEDGNILGHKLEYNYNSHNVYITADKQKNGIKANFKIRHNIEDGGVQLADHYQQNTPIGDGPVLLPDNHYLSYQSALSKDPNEKRDHMVLLEFVTAAGITLGMDELYK |
| mGFPmut3 | SKGEELFTGVVPILVELDGDVNGHKFSVSGEGEGDATYGKLTLKFICTTGKLPVPWPTLVTTFGYGVQCFARYPDHMKQHDFFKSAMPEGYVQERTIFFKDDGNYKTRAEVKFEGDTLVNRIELKGIDFKEDGNILGHKLEYNYNSHNVYIMADKQKNGIKVNFKIRHNIEDGSVQLADHYQQNTPIGDGPVLLPDNHYLSTQSKLSKDPNEKRDHMVLLEFVTAAGITHGMDELYK |
| meGFP | SGGGGSKVSKGEELFTGVVPILVELDGDVNGHKFSVSGEGEGDATYGKLTLKFICTTGKLPVPWPTLVTTLTYGVQCFSRYPDHMKQHDFFKSAMPEGYVQERTIFFKDDGNYKTRAEVKFEGDTLVNRIELKGIDFKEDGNILGHKLEYNYNSHNVYIMADKQKNGIKVNFKIRHNIEDGSVQLADHYQQNTPIGDGPVLLPDNHYLSTQSKLSKDPNEKRDHMVLLEFVTAAGITLGMDELYK |
| Dronpa | VIKPDMKIKLRMEGAVNGHPFAIEGVGLGKPFEGKQSMDLKVKEGGPLPFAYDILTTVFCYGNRVFAKYPENIVDYFKQSFPEGYSWERSMNYEDGGICNATNDITLDGDCYIYEIRFDGVNFPANGPVMQKRTVKWEPSTEKLYVRDGVLKGDVNMALSLEGGGHYRCDFKTTYKAKKVVQLPDYHFVDHHIEIKSHDKDYSNVNLHEHAEAHSELPRQAK |
| Dendra2 | MNTPGINLIKEDMRVKVHMEGNVNGHAFVIEGEGKGKPYEGTQTANLTVKEGAPLPFSYDILTTAVHYGNRVFTKYPEDIPDYFKQSFPEGYSWERTMTFEDKGICTIRSDISLEGDCFFQNVRFKGTNFPPNGPVMQKKTLKWEPSTEKLHVRDGLLVGNINMALLLEGGGHYLCDFKTTYKAKKVVQLPDAHFVDHRIEILGNDSDYNKVKLYEHAVARYSPLPSQVW |
| E2-Crimson | DSTENVIKPFMRFKVHMEGSVNGHEFEIEGVGEGKPYEGTQTAKLQVTKGGPLPFAWDILSPQFFYGSKAYIKHPADIPDYLQSFPEGFKWERVMNFEDGGVVTVTQDSSLQDGTLIYHVKFIGVNFPSDGPVMQKKTLGWEPSTERNYPRDGVLKGENHMALKLKGGGHYLCEFKSIYMAKKPVKLPGYHYVDYKLDITSHNEDYTVVEQYERAEARHHLFQ |
| dsRed | RSSKNVIKEFMRFKVRMEGTVNGHEFEIEGEGEGRPYEGHNTVKLKVTKGGPLPFAWDILSPQFQYGSKVYVKHPADIPDYKKLSFPEGFKWERVMNFEDGGVVTVTQDSSLQDGCFIYKVKFIGVNFPSDGPVMQKKTMGWEASTERLYPRDGVLKGEIHKALKLKDGGHYLVEFKSIYMAKKPVQLPGYYYVDSKLDITSHNEDYTIVEQYERTEGRHHLFL |

**Table S1.** The amino acid sequences of MreB and the different fluorescent proteins used in this study. Linker amino acid sequences are shown in yellow and the location of the fluorescent protein is in red.

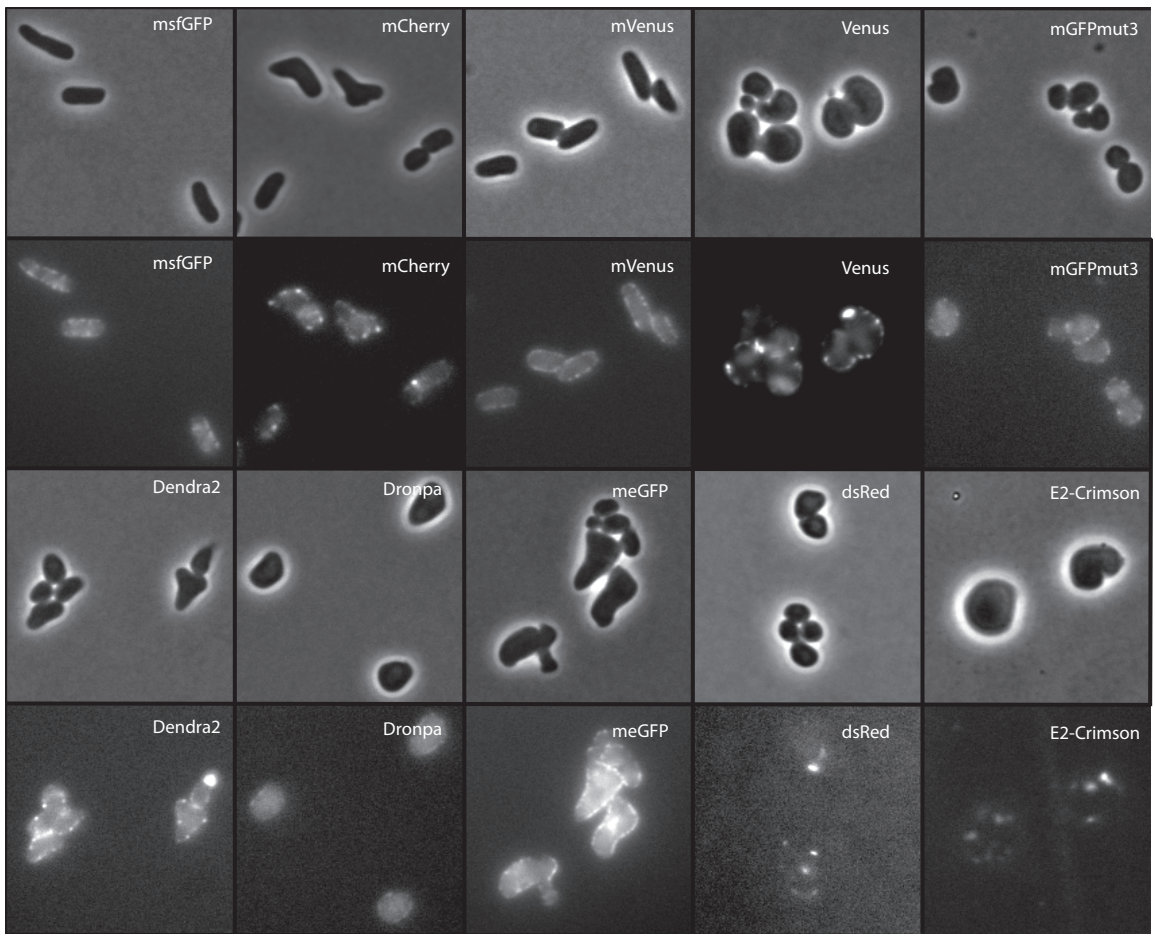

**Figure S1**. The MG1655 *mreB* deletion strain was complemented with a plasmid containing the MreB operon in which MreB is labeled with different fluorescent fusions to MreB. The florescent proteins that have been previously shown to cause the least amount of dimerization are msfGFP, mVenus, mGFPmut3, Dendra 2, Dronpa, and meGFP. msfGFP was best able to complement rod shape and had optimal quantum yield for prolonged imaging. Venus is known to form dimers and both dsRed and E2-Crimson form tetramers. Each fusion is imaged using phase microscopy (top row) and fluorescence microscopy (bottom row).

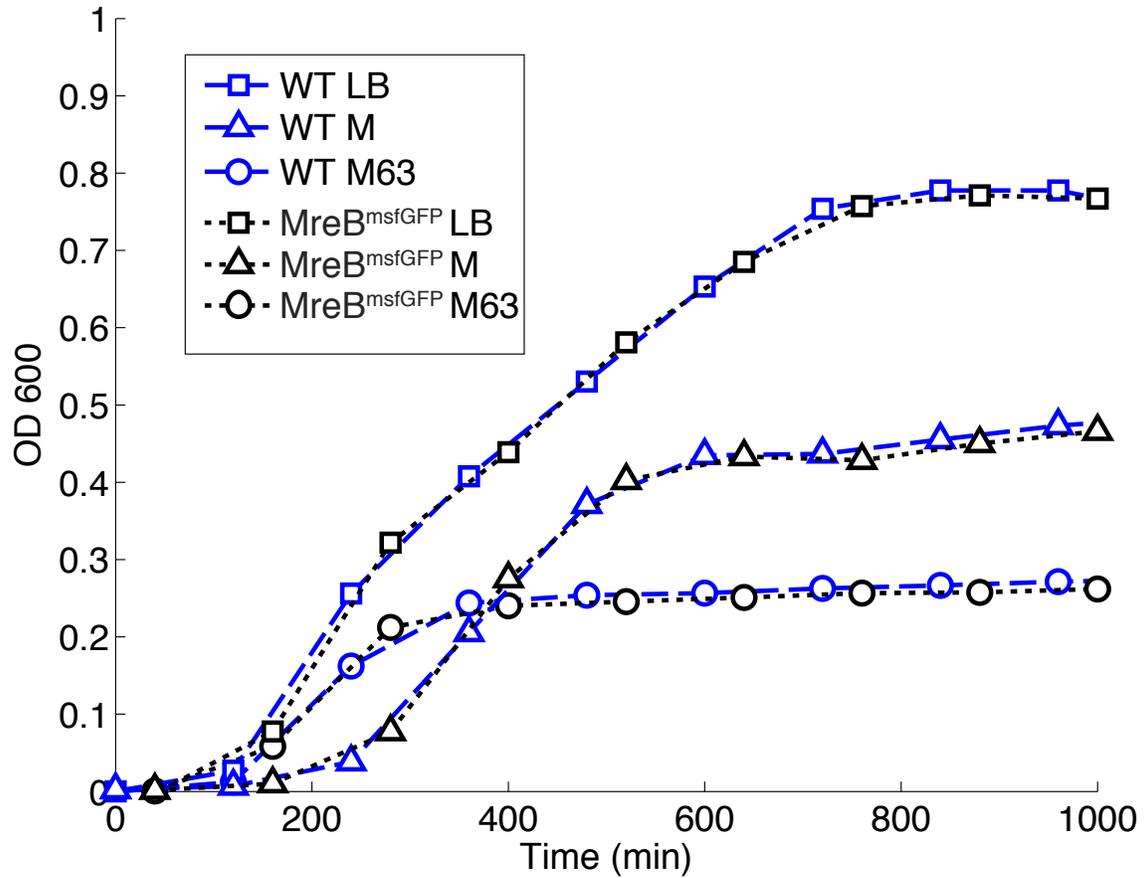

**Figure S2.** Comparison of OD growth curves between *E. coli* expressing native MreB and *E. coli* expressing tagged MreB$^{msfGFP}$ integrated in the native *mreB* locus. Cells were grown in LB, M63 media with glucose and casamino acids, and in M media. There is close agreement between the two strains in all types of media. Data was averaged over 3 replicates.

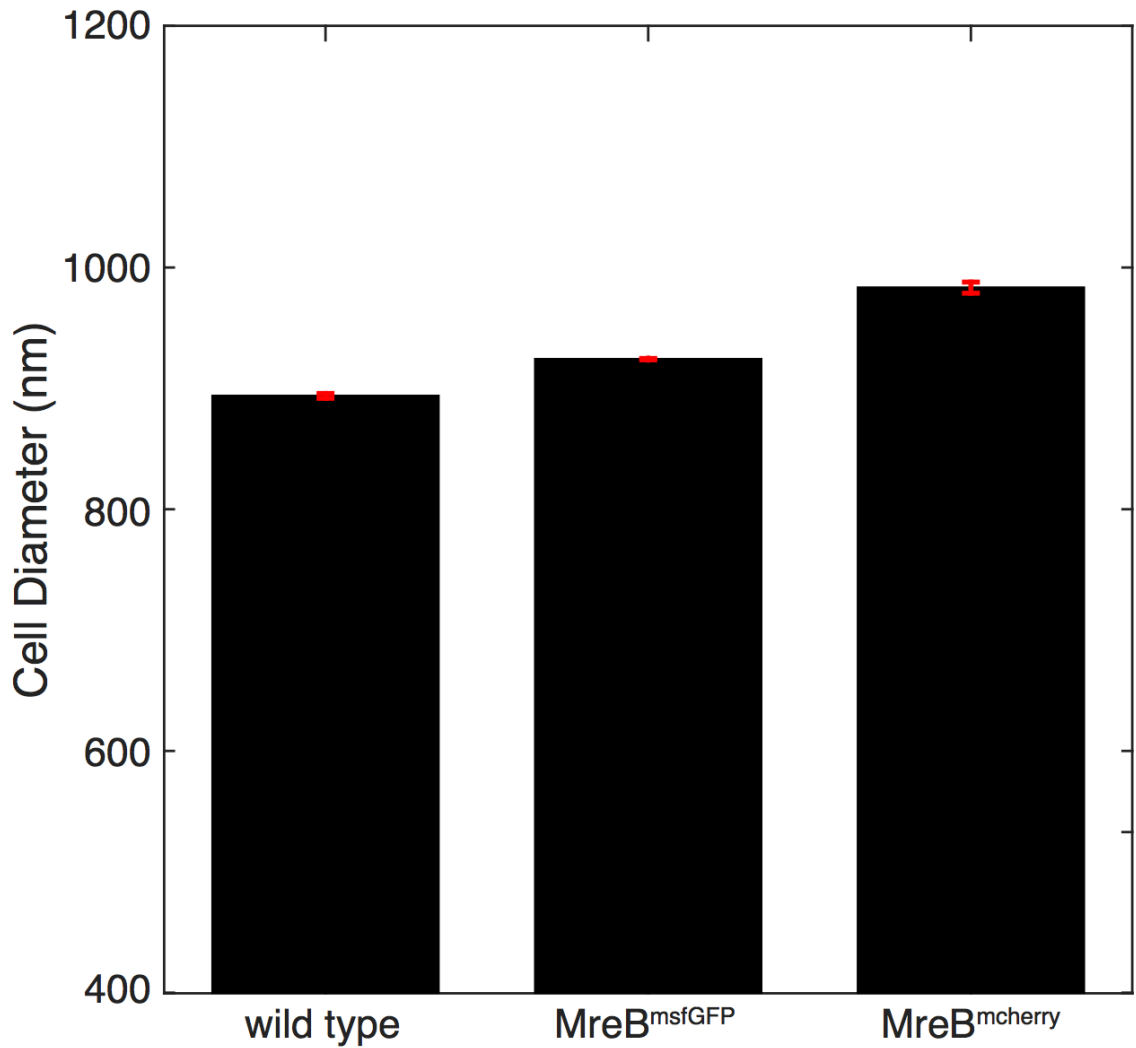

**Figure S3.** Average cell diameters for *E. coli* expressing native MreB (n=645), MreB$^{msfGFP}$ (n= 459), and MreB$^{mcherry}$ (n=372). The average diameters were 893±3 nm for the unlabeled strain, 934±6 nm for MreB$^{msfGFP}$, and 983 ±5 nm for MreB$^{mcherry}$. Cells were grown in M63 media with casamino acids.

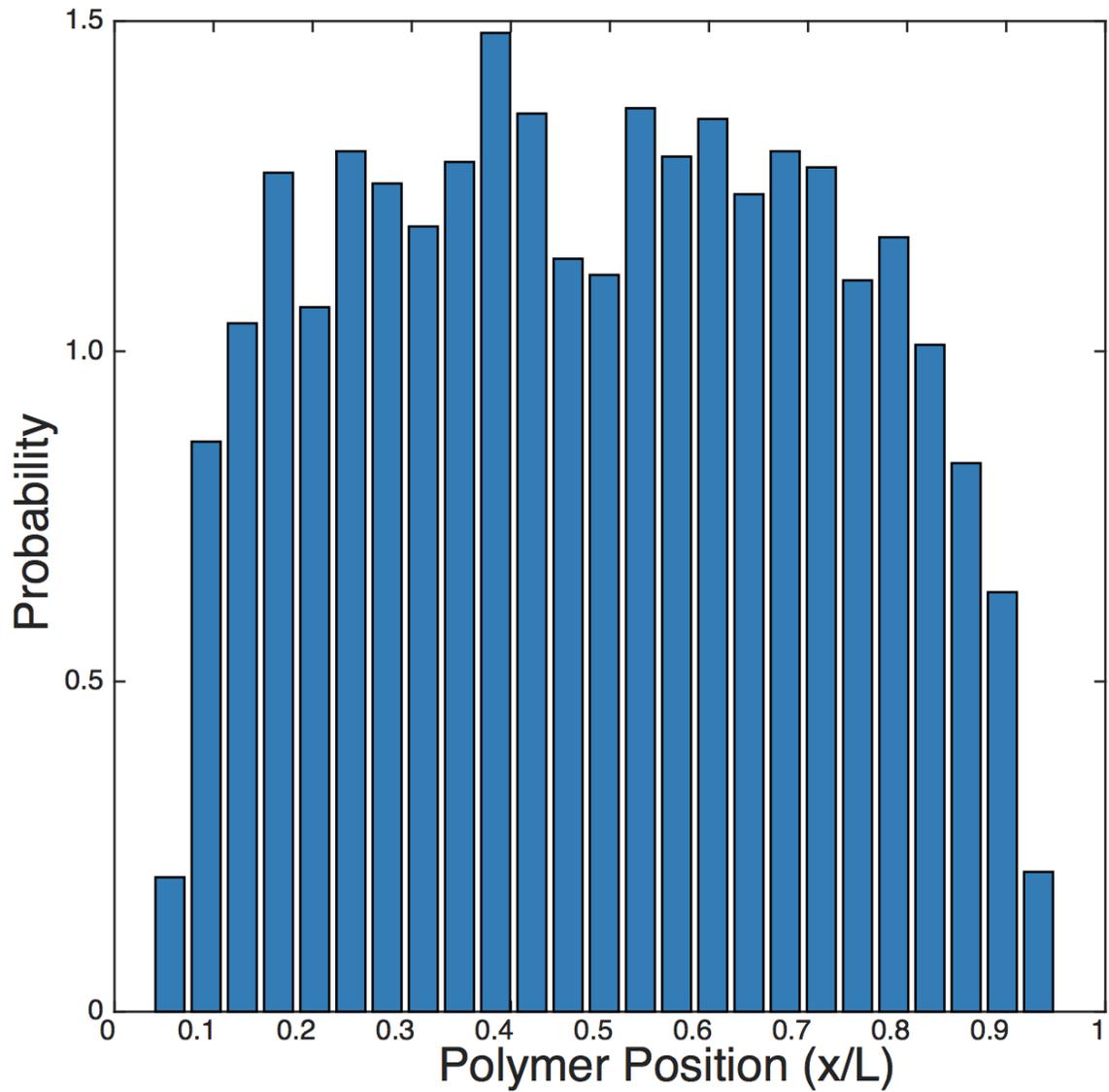

**Figure S4.** A distribution of MreB polymer positions as a function of percentage length along the cell in *E. coli* expressing MreB$^{msfGFP}$. The polymers are excluded near the poles of the cells. Data is collected from 459 cells, with an average of 7.3 polymers detected per cell.

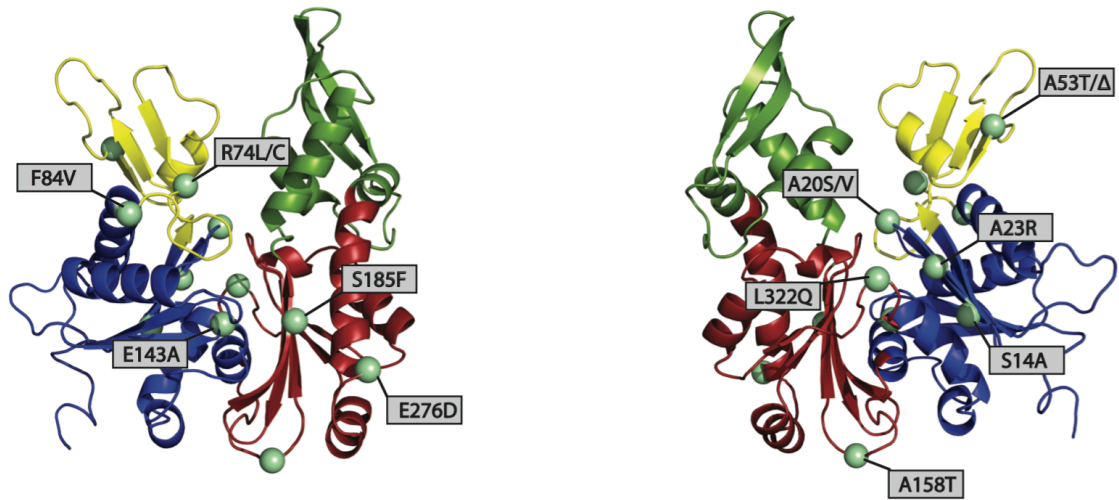

**Figure S5.** MreB amino acid substitutions are found spanning subdomains IA (Blue), IB (Yellow), and IIA (Red)(1). Some residues are hit more than once. *E. coli* MreB structure was generated using the Phyre2 server (2).

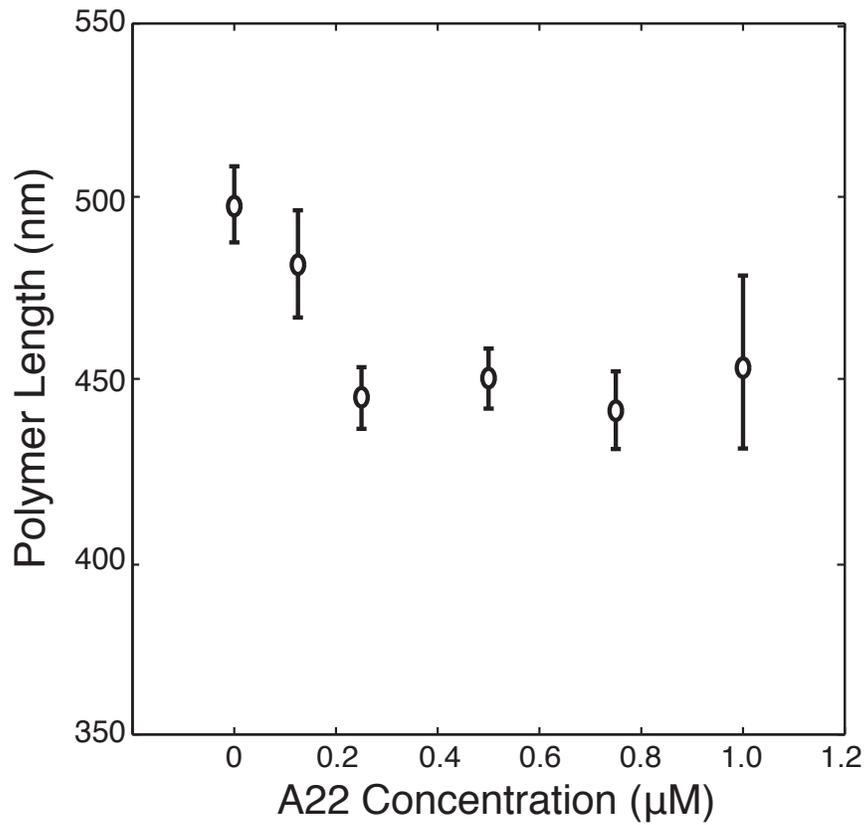

**Figure S6.** MreB polymer length as a function of A22 concentration for cells expressing MreB$^{msfGFP}$. Cells were grown in the presence different sub-lethal concentrations of the MreB polymerization inhibitor A22 for multiple generations and imaged in exponential growth phase. Error bars indicate 80% confidence intervals.

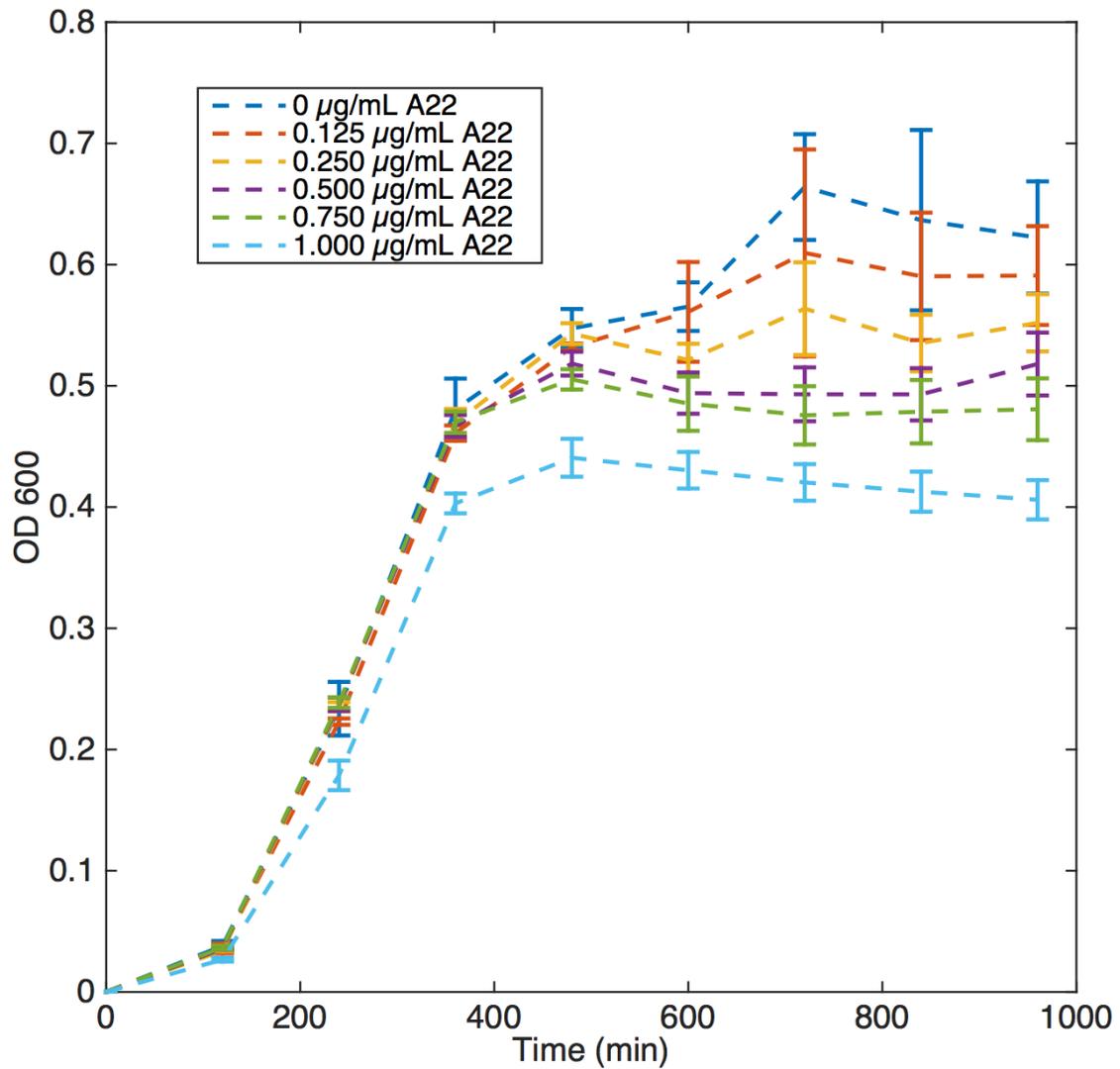

**Figure S7.** OD600 growth curves for *E. coli* grown at different sub-lethal concentrations of the MreB polymerization inhibitor A22. Cells grown at higher concentrations of A22 have lower log phase growth rates and lower steady state OD.

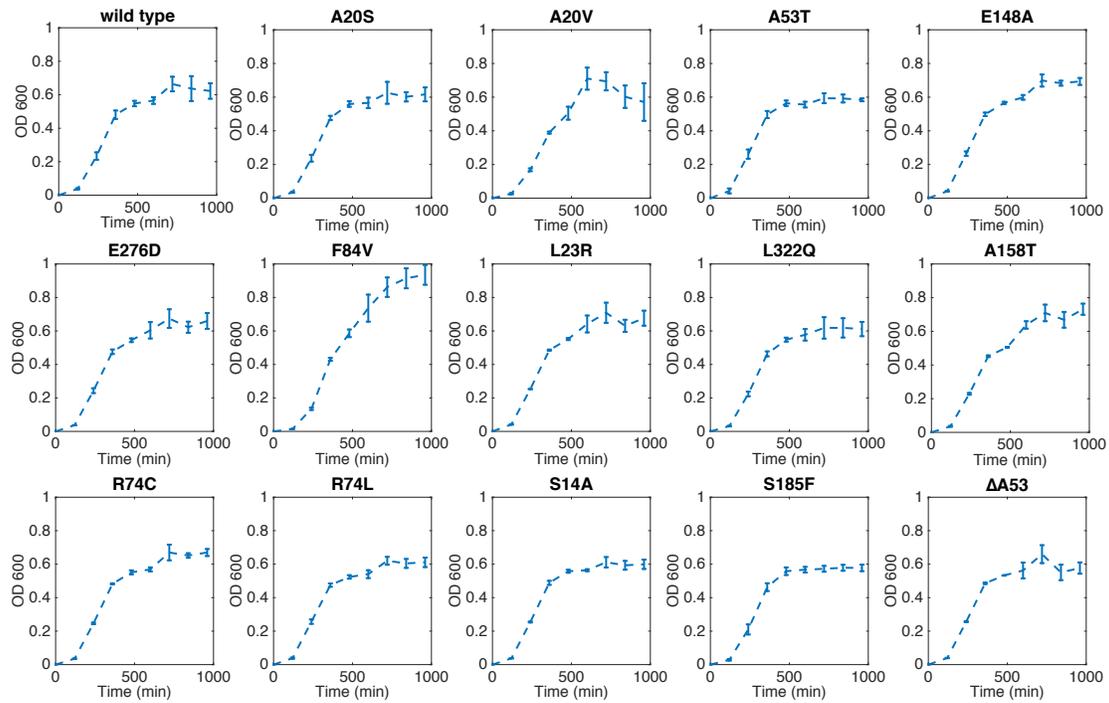

**Figure S8.** OD600 growth curves for the *E. coli* MreB mutants used in this study. All mutants except F84V have comparable growth rates and steady state ODs. F84V shows the slowest growth rate yet reaches the highest final OD.